\documentclass{PHYEAUTH}
\usepackage{graphicx}
\usepackage{amsmath}
\usepackage{amssymb}
\usepackage[usenames]{color}

\def\stacksymbols #1#2#3#4{\def\theguybelow{#2}
    \def\verticalposition{\lower#3pt}
    \def\spacingwithinsymbol{\baselineskip0pt\lineskip#4pt}
    \mathrel{\mathpalette\intermediary#1}}
\def\intermediary#1#2{\verticalposition\vbox{\spacingwithinsymbol
      \everycr={}\tabskip0pt
      \halign{$\mathsurround0pt#1\hfil##\hfil$\crcr#2\crcr
               \theguybelow\crcr}}}
\def\lapproxeq{\stacksymbols{<}{\sim}{2.5}{.2}}

\begin{document}

\begin{frontmatter}

\title{Corner Multifractality for Reflex Angles and Conformal Invariance
at 2D Anderson Metal-Insulator Transition with Spin-Orbit Scattering}

\author[address1]{H.\ Obuse},
\author[address2]{A.\ R.\ Subramaniam},
\author[address1]{A.\ Furusaki},
\author[address2]{I.\ A.\ Gruzberg},
\author[address3]{A.\ W.\ W.\ Ludwig}

\address[address1]{Condensed Matter Theory Laboratory, RIKEN, Wako,
 Saitama 351-0198, Japan}
\address[address2]{James Franck Institute, University of Chicago, 5640
 South Ellis Avenue, Chicago, Illinois 60637, USA}
\address[address3]{Physics Department, University of California, Santa
 Barbara, California 93106, USA}

\begin{abstract}

We investigate boundary multifractality of critical wave
functions at the Anderson metal-insulator transition in two-dimensional
disordered non-interacting electron systems with spin-orbit scattering.
We show numerically that multifractal exponents at a corner with an opening
angle $\theta=3\pi/2$ are directly related to those near a straight
boundary in the way dictated by conformal symmetry.
This result extends our previous numerical results on corner multifractality
obtained for $\theta<\pi$ to $\theta > \pi$, and gives further supporting
evidence for conformal invariance at criticality.
We also propose a refinement of the validity of the symmetry relation
of A. D. Mirlin \textit{et al.}, Phys.\ Rev.\ Lett.\ \textbf{97} (2006)
046803, for corners.

\end{abstract}


\end{frontmatter}

Anderson metal-insulator transitions are continuous phase transitions
driven by disorder.
Examples of localization-delocalization  (Anderson) transitions occurring
in two-dimensions (2D) include non-interacting electronic systems with
spin-orbit scattering (`symplectic symmetry class'),
with sublattice symmetry, or in strong magnetic fields (quantum Hall effect).

Recently, we have reported numerical evidence for the presence of
conformal invariance at the 2D Anderson transition in the symplectic
symmetry class \cite{Obuse2007}.
To that end, we have considered multifractal properties of critical wave
functions near boundaries of disordered samples of finite size, and
verified numerically that the multifractal exponents of critical wave
functions at corners with opening angle $\theta$ (corner multifractality)
are related, through simple relations derived from conformal invariance,
to the exponents computed near straight edges (surface multifractality).
In Ref.~\cite{Obuse2007} we have discussed corner multifractality at
wedges with angles $\theta<\pi$ only, both acute ($\theta = \pi/4$) and
obtuse ($\theta = 3\pi/4$).
In this paper we extend this analysis to a corner with $\theta=3\pi/2$
to show that the same equation relating surface and corner multifractality
holds for the corner with a reflex angle ($\theta > \pi$).

Following Refs.~\cite{Obuse2007,Subramaniam2006}, we define bulk, surface,
and corner multifractality from the scaling of moments of wave functions
$\psi(\boldsymbol{r})$ in bulk (b), surface (s), and corner ($\theta$)
regions,
\begin{align}
L^{d_\mathrm{x}} \overline{ |\psi(\boldsymbol{r})|^{2q}} &\sim
L^{-\tau_q^\mathrm{x}}, & (\mathrm{x}=\theta, \mathrm{s},
\mathrm{b}),
\label{eq:tau_q}
\end{align}
where $d_\mathrm{x}$ is the spatial dimension of each region
($d_\mathrm{b}=2$, $d_\mathrm{s}=1$, and $d_\theta=0$).
The overbar represents the ensemble (disorder) average and the simultaneous
spatial average over a region x surrounding the point $\boldsymbol{r}$.
The exponents $\tau_q^\mathrm{b}$, $\tau_q^\mathrm{s}$, and $\tau_q^\theta$
are the bulk, surface, and corner multifractal exponents, respectively.
From the multifractal exponents we extract non-vanishing anomalous
dimensions $\Delta_q^\mathrm{x}$,
\begin{align}
\Delta_q^\mathrm{x} = \tau_q^{\mathrm{x}} - 2q + d_\mathrm{x}.
\label{eq:Delta_q}
\end{align}
The multifractal singularity spectra $f^\mathrm{x}(\alpha)$ are obtained
from $\tau^\mathrm{x}_q$ by Legendre transformation,
\begin{align}
f^\mathrm{x}(\alpha^\mathrm{x}) &= \alpha^\mathrm{x} q -
\tau^\mathrm{x}_q, & \alpha^\mathrm{x} = \frac{d\tau^\mathrm{x}_q}{dq}.
\label{eq:Legendre}
\end{align}

As explained in Ref.~\cite{Obuse2007}, under the assumption that the $q$-th
moment $\overline{|\psi(\boldsymbol{r})|^{2q}}$ is represented by a primary
operator in an underlying conformal field theory \cite{Belavin1984},
one can derive, using the conformal mapping $w=z^{\theta/\pi}$,
the relation between the surface and corner multifractal spectra
$f^\mathrm{x}(\alpha^\mathrm{x}_q)$,
\begin{equation}
\alpha_q^{\theta}-2 = \frac{\pi}{\theta}(\alpha_q^\mathrm{s}-2), \quad
f^\theta(\alpha_q^\theta) = \frac{\pi}{\theta}
\left[ f^\mathrm{s}(\alpha_q^{\text{s}}) - 1 \right ].
\label{eq:alpha^theta}
\end{equation}
The validity of these relations provides direct evidence for conformal
invariance at a 2D Anderson transition and for the primary
nature of the operator.

In Ref.~\cite{Obuse2007} we have shown that the probability distribution
of $\ln|\psi(\boldsymbol{r})|^2$ becomes broader, as the opening angle
$\theta$ is reduced.
This implies that the distribution is narrower at a corner
with larger $\theta$.
We may thus expect that multifractal exponents can be more accurately
calculated for corners with reflex angles than for corners
with angles $\theta < \pi$.
We can then estimate the surface $f^\mathrm{s}(\alpha_q^\mathrm{s})$
by taking the  numerical data for $\alpha_q^\theta$ and
$f^\theta(\alpha_q^\theta)$ obtained for $\theta>\pi$ as input into
Eq.\ (\ref{eq:alpha^theta}).
Moreover, we can relate multifractal spectra of corners with different
angles $\theta$ and $\theta'$ ($\theta<\theta'$), yielding
\begin{equation}
\alpha_q^{\theta}-2 = \frac{\theta'}{\theta}(\alpha_q^{\theta'}-2), \quad
f^\theta(\alpha_q^\theta) = \frac{\theta'}{\theta}
f^{\theta'}(\alpha_q^{\theta'}).
\label{eq:alpha^theta_corner}
\end{equation}
As we pointed out in Ref.\ \cite{Obuse2007}, Eqs.\ (\ref{eq:alpha^theta})
and (\ref{eq:alpha^theta_corner}) are valid only if all occurring
$\alpha_q^{\mathrm{x}} \geqslant 0$, because $\alpha_q^{\mathrm{x}}$ is
non-negative for normalized wave functions.
Thus, when the prefactor $\theta'/\theta$ is larger than one
(and hence $0 \leqslant \alpha^\theta_q < \alpha^{\theta'}_q$),
the first of Eq.~(\ref{eq:alpha^theta_corner}) cannot be used
for $q>q_\theta^{}$, where $q_\theta^{}$ is a solution to
$\alpha_q^\theta=0$ in Eq.~(\ref{eq:alpha^theta_corner}).
(We do not know if $q_\theta^{}$ is finite for $\theta \geqslant \pi$.)
Taking this physical constraint into account, we find the following
relation between anomalous dimensions for corners ($\theta<\theta'$),
\begin{equation}
\Delta_q^\theta = \left\{
\begin{aligned}
& \dfrac{\theta'}{\theta} \Delta_{q}^{\theta'}, & q \leqslant q_\theta^{}, \\
& \dfrac{\theta'}{\theta} \Delta_{q_\theta^{}}^{\theta'}-2(q-q_\theta^{}),
& q > q_\theta^{}.
\end{aligned}
\right. \label{eq:Delta_q^theta_corner}
\end{equation}
If we set $\theta=\pi$ in Eq.\ (\ref{eq:Delta_q^theta_corner}), we obtain
a relation between anomalous dimensions at a surface ($\theta=\pi$) and
a corner with a reflex angle ($\theta'>\pi$).
In Ref.\ \cite{Obuse2007} we also discussed the symmetry relation of
Ref.\ \cite{mirlin}, $\Delta^\mathrm{x}_q = \Delta^\mathrm{x}_{1-q}$,
and its application to corners $\mathrm{x}=\theta$.
Here we propose, as a refinement of that discussion, that this symmetry
relation (i) is valid for corners of any angle $\theta$ including
$\theta=\pi$ (straight boundaries), but only in the range of $q$ satisfying
$1 - q_\theta^{} \leqslant q \leqslant q_\theta^{}$, corresponding
precisely \cite{Obuse2007,mirlin} to the range
$0 \leqslant \alpha^\theta_q \leqslant 4$, and (ii) makes no statements
about $\Delta^\theta_q$ for values of $q$ outside of this range.
[The dependence on $q$ of $\Delta^\theta_q$ is linear for $q > q_\theta^{}$
(corresponding to the termination of the multifractal
spectrum \cite{MirlinEtAl-RMP}), whilst it may, in general, continue to
be non-linear for $q < 1-q_\theta^{}$, even \cite{Obuse2007,mirlin} when
$\alpha^\theta_q > 4$.]

\begin{figure}[b]
\centering
\includegraphics[width=0.99\linewidth]{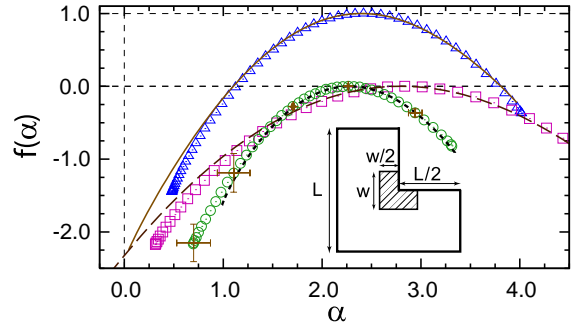}
\caption{(color online) Multifractal spectra $f(\alpha)$ for corners
with $\theta=3\pi/2$ (circles) and $\theta=\pi/2$ (squares), and surface
(triangles) regions.
Error bars are plotted at integer values of $q$ for the corner
with $\theta=3\pi/2$.
The solid and short-dashed curves represent the theoretical prediction
from Eq.\ (\ref{eq:alpha^theta}), where $f^{3\pi/2}(\alpha_q^{3\pi/2})$
and $f^{\mathrm{s}}(\alpha_q^{\mathrm{s}})$ are used as input,
respectively.
The dashed curve is calculated from Eq.\ (\ref{eq:alpha^theta_corner})
with $f^{3\pi/2}(\alpha_q^{3\pi/2})$ used as input.
Inset: L-form geometry with $3L^2/4$ sites.
The shaded part is the corner region with $\theta=3\pi/2$ of
the size $3w^2/4$.}
\label{fig:f(alpha)}
\end{figure}

In this work, we numerically verify these relations by computing corner
multifractal spectra at $\theta=3\pi/2$ for the L-shape samples shown
in the inset of Fig.~\ref{fig:f(alpha)}.
We take a tight-binding model with both random on-site potential and
random SU(2) hopping \cite{Asada2002}, and numerically obtain, with the
forced oscillator method \cite{Nakayama2001}, a wave function $\psi$
having energy eigenvalue closest to a critical point $E_c=1.0$ (in units
of the mean hopping) for each random realization characterized by the
on-site disorder strength $W_c=5.952$.
The system size $L$ is varied through $L=24,30, \cdots, 120$ and the
number of disordered samples is $6\times10^4$ for each $L$.
We set $w=2$ of the corner region shown in Fig.~\ref{fig:f(alpha)}.
Multifractal spectra are computed in the same way as in Ref.~\cite{Obuse2007}.

In Fig.~\ref{fig:f(alpha)}, we show multifractal spectra $f(\alpha)$ of
corners with $\theta=3\pi/2$, together with those of corners with
$\theta=\pi/2$, and of the  surface region \cite{Obuse2007}.
The peak position $\alpha_0^{\mathrm{x}}$ of $f^{3\pi/2}(\alpha_q^{3\pi/2})$
is $\alpha_0^{3\pi/2}=2.265 \pm 0.003$, which is smaller than
$\alpha_0^{\pi/2}=2.837 \pm 0.003$.
Also, the width of $f^{3\pi/2}(\alpha)$ is smaller than that of
$f^\mathrm{s}(\alpha_q^\mathrm{s})$.
This is consistent with Eq.~(\ref{eq:alpha^theta}) at $\theta=3\pi/2$.
Figure \ref{fig:f(alpha)} clearly shows that $f^{3\pi/2}(\alpha)$ computed
directly for the corner with $\theta=3\pi/2$ agrees well with the
short-dashed curve obtained from Eq.~(\ref{eq:alpha^theta}) while using
$f^\mathrm{s}(\alpha)$ as input, which verifies Eq.~(\ref{eq:alpha^theta})
derived from conformal invariance.
We have also calculated the surface $f^{\mathrm{s}}(\alpha_q^{\mathrm{s}})$
and corner $f^{\pi/2}(\alpha_q^{\pi/2})$ from Eqs.~(\ref{eq:alpha^theta})
and (\ref{eq:alpha^theta_corner}), respectively, using
$f^{3\pi/2}(\alpha_q^{3\pi/2})$ as input into these equations.
This allows us to estimate $f^{\mathrm{s}}(\alpha_q^{\mathrm{s}})$
near $\alpha^\mathrm{s}\approx0$, providing an estimate for
$q_\mathrm{s}^{} = d f^\mathrm{s}(\alpha^\mathrm{s}=0)/d \alpha^\mathrm{s}$.
The theoretical predictions (solid and dashed curves) are in good agreement
with the numerical data (triangles and squares) for
$f^\mathrm{s}(\alpha_q^{\mathrm{s}})$ and
$f^{\pi/2}(\alpha_q^{\pi/2})$, respectively.

\begin{figure}[t]
\centering
\includegraphics[width=\linewidth]{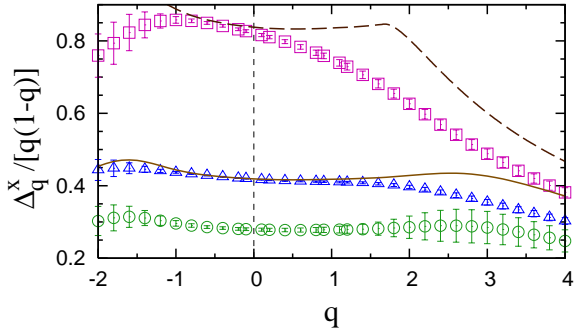}
\caption{(color online) The exponents $\Delta_q^\mathrm{x}/[q(1-q)]$ for
corner with $\theta=3\pi/2$ (circles) and $\theta=\pi/2$ (squares),
and surface (triangles) regions.
Solid and dashed curves represent the conformal relation
(\ref{eq:Delta_q^theta_corner}).}
\label{fig:Delta}
\end{figure}

Figure \ref{fig:Delta} shows the anomalous dimensions
$\Delta_q^{\mathrm{x}}$ for corners with $\theta=3\pi/2$ and
$\theta=\pi/2$, and the surface region, which are numerically calculated
from the scaling
$\overline{|\psi(\boldsymbol{r})|^{2q}}/
(\overline{|\psi(\boldsymbol{r})|^2})^q
\sim L^{-\Delta_q^{\mathrm{x}}}$.
The solid and dashed curves represent the theoretical prediction,
Eq.~(\ref{eq:Delta_q^theta_corner}), from the conformal mapping
using $\Delta_q^{3\pi/2}$ as inputs.
The data points for $\Delta_q^\mathrm{s}$ (triangles) agree with the
solid curve for $|q| \lapproxeq 1.5$, while those for $\Delta_q^{\pi/2}$
(squares) are close to the dashed curve only near $q \approx 0$.
The data points for $\Delta_q^{3\pi/2}$ satisfy, within error bars,
the symmetry relation \cite{mirlin}
$\Delta_q^{3\pi/2}=\Delta_{1-q}^{3\pi/2}$ in the vicinity of $q=1/2$,
indicating good numerical accuracy.
This opens the possibility that one can use corner multifractality
for $\theta>\pi$ to obtain, with the help of Eqs.~(\ref{eq:alpha^theta})
and (\ref{eq:Delta_q^theta_corner}), more accurate estimates for
multifractal properties at a straight surface and corners with $\theta<\pi$.

We briefly comment on multifractality of a whole sample with boundaries.
We have found in Refs.~\cite{Obuse2007,Subramaniam2006} that corner
multifractality may dominate multifractality of a whole system, even in
the thermodynamic limit, for large values of $|q|$ if
$\tau_q^{\theta}< \tau_q^{\mathrm{b}}, \tau_q^{\mathrm{s}}$.
Here we point out that this cannot happen with corners of reflex angles
($\theta>\pi$).
The proof goes as follows.
We first note that, from Eqs.~(\ref{eq:Delta_q}) and
(\ref{eq:Delta_q^theta_corner}), the difference of corner and surface
multifractal exponents is given by
$\tau_q^\theta - \tau_q^{\mathrm{s}} =
1 + \Delta_q^{\mathrm{s}} (\pi/\theta-1)$ as long as
$\alpha_q^\mathrm{s} > 0$
(note that $\alpha^\theta_q > \alpha^\mathrm{s}_q$  for $\theta > \pi$).
Thus, when $\pi < \theta < 2\pi$, the inequality
$\tau_q^\theta > \tau_q^\mathrm{s}$ holds if
$\Delta_q^\mathrm{s} \leqslant 2$.
Secondly, since $\tau_q^\mathrm{x}$ is a convex function of $q$ with
the constraints $\tau_0^{\mathrm{x}}=-d_{\mathrm{x}}$ and
$\tau_1^{\mathrm{x}}=2-d_{\mathrm{x}}$ (recalling \cite{Obuse2007} $\mu=0$,
and thus $\Delta^\mathrm{x}_1=0$), we find
$\Delta_q^{\mathrm{x}} \leqslant 0$ for $|q-1/2| \geqslant 1/2$ and
$0 < \Delta_q^{\mathrm{x}} < 2$ for $0 < q < 1$.
We thus conclude that $\tau_q^\theta > \tau_q^\mathrm{s}$ when
$\alpha_q^\mathrm{s} > 0$.
Finally, when $\tau_q^\theta - \tau_q^{\mathrm{s}}$ is positive for
$\alpha_q^{\mathrm{s}}>0$ the difference remains positive even if
$q_\mathrm{s}^{} <\infty$ and in the regime $q\geqslant q_\mathrm{s}^{}$
where $\alpha_{q_\mathrm{s}}^{\mathrm{s}}=0$, because $\tau_q^{\mathrm{s}}$
is then constant for $q \geqslant q_\mathrm{s}^{}$
(and $d \tau^\theta_q/dq = \alpha^\theta_q \geqslant \alpha^\mathrm{s}_q$).
Hence, contributions from corner multifractality at $\theta>\pi$ cannot
be larger than contributions from surface multifractality.
The numerical results shown in Fig.~\ref{fig:f(alpha)} are consistent
with the above general argument.

In summary, we have investigated corner multifractality for the reflex
angle $\theta=3\pi/2$ and confirmed the validity of the conformal
symmetry relations.
This result provides stronger evidence for the presence of conformal
symmetry at the 2D Anderson metal-insulator transition with spin-orbit
scattering.

This work was supported by NAREGI grant from MEXT of Japan, the NSF under
Grant No.\ PHY99-07949, the NSF MRSEC under No.\ DMR-0213745, the NSF Career
Grant No.\ DMR-0448820, and Research Corporation. Numerical calculations
were performed on the RIKEN Super Combined Cluster System.
%

\end{document}